# (c) Joan Masso
#
\documentstyle[preprint,aps,prd]{revtex}
\tighten
\begin{document}
\draft

\title{First order hyperbolic formalism for Numerical Relativity}
\author{$^{(1)}$C.~Bona, $^{(2,3)}$J.~Mass\'o, $^{(2,3,4)}$E.~Seidel and
$^{(1)}$J.~Stela}

\address{
$^{(1)}$Departament de Fisica, Universitat de les Illes Balears\\
$^{(2)}$Max-Planck-Institut f{\"u}r Gravitationsphysik,
Schlaatzweg 1, 14473 Potsdam, Germany \\
$^{(3)}$ National Center for Supercomputing Applications,
Beckman Institute, 405 N. Mathews Ave., Urbana, IL, 61801 \\
$^{(4)}$ Departments of Physics and Astronomy,
University of Illinois, Urbana, IL 61801 \\
}

\date{23 March 1997}

\maketitle
\begin{abstract}
The causal structure of Einstein's evolution equations is considered.
We show that in general they can be written as a first order system of
balance laws for {\em any} choice of slicing or shift.  We also show
how certain terms in the evolution equations, that can lead to
numerical inaccuracies, can be eliminated by using the Hamiltonian
constraint.  Furthermore, we show that the entire system is hyperbolic
when the time coordinate is chosen in an invariant algebraic way, and
for any fixed choice of the shift.  This is achieved by using the
momentum constraints in such as way that no additional space or time
derivatives of the equations need to be computed.  The slicings that
allow hyperbolicity in this formulation belong to a large class,
including harmonic, maximal, and many others that have been commonly
used in numerical relativity.  We provide details of some of the
advanced numerical methods that this formulation of the equations
allows, and we also discuss certain advantages that a hyperbolic
formulation provides when treating boundary conditions.
\end{abstract}

\section{Introduction and Overview}
In a previous Letter~\cite{PRL95}, we proposed a new formalism for
numerical relativity based on a formulation of Einstein Field
Equations as a hyperbolic system of balance laws.  This was an
extension of previous results which were derived originally in a
particular gauge (harmonic slicing) ~\cite{PRL92}, but in
~\cite{PRL95} we showed how to enlarge this to a broad family of
slicing conditions, including the most commonly used choices in
numerical relativity.  The application of this formalism to practical
problems requires a more detailed description and discussion, which is
the aim of the present work.  In this first follow-up paper we describe
the formalism in much more detail than before, and describe broadly the kinds
of numerical methods that are applicable to such a system of
equations.  In future papers in this series we will provide detailed
numerical examples in 1D and 3D, with comparisons to other formalisms
and standard numerical methods.

There are numerous motivations for this new formulation of the
equations:

{\em (i) Numerical Methods.} First, standard numerical methods for
evolution systems, such as flux conservative balance laws, have been
developed specifically to treat only certain systems of equations, and
only for these systems are their numerical properties well understood.
The standard ADM formulation of the equations~\cite{Lich,Choquet,ADM} is much
more complicated, and therefore one has to be very careful when attempting
to apply {\em ad hoc} variations on these methods to Einstein's
theory; the numerical properties of these systems are not well
understood.  This is a major reason why numerical relativity has
proved much more difficult than, say, computational fluid dynamics
(CFD).  For example, without novel approaches, such as apparent
horizon boundary conditions~\cite{Seidel92a,Anninos94e}, black hole
spacetimes, could not previously be evolved beyond about
$t=150M$~\cite{Bernstein89,Shapiro86} without codes crashing due to
the inadequacy of the numerical methods being used.  Worse yet, in
interesting cases where black holes (i.e.  horizons) do {\em not} seem
to form, yet where singularities may be developing, similar problems
cause codes to become very inaccurate and crash even much
earlier~\cite{Shapiro91}, preventing a full exploration of the
spacetime.  If special numerical numerical methods could be developed
specifically for the standard formulation of the Einstein equations, as
they have been for decades in CFD, presumably these problems could be
treated properly.  However, this would be a formidable undertaking.
On the other hand we have managed to write the equations in a form
which can take advantage of this vast knowledge of numerical methods
applied to systems of conservations laws, and their numerical
properties, developed for CFD. With this new formulation one can now
apply many standard methods (e.g.  the genuine
MacCormack~\cite{maccormack} method) for {\em any} choice of lapse and
shift.  This already offers new possibilities for evolution schemes.
But for a wide family of lapse conditions, the system is furthermore
{\em hyperbolic}, meaning here that one can find a complete set of
eigenfields with real eigenvalues for the system of equations.  This
provides a framework for developing a much deeper knowledge of the
system of equations, its characteristic fields and their speeds, and
also brings a variety of more advanced numerical methods (e.g., so-called TVD
schemes~\cite{TVD}) that exploit this knowledge at the finite
difference level.  In CFD this knowledge is crucial in treating the
kinds of large gradients and ``shock-like'' features that can also
develop in strongly gravitating systems due to gauge or physical
effects in the Einstein system.  We will give examples of these
possibilities below.

{\em (ii) Boundary conditions.} Second, in the case of slicings that
allow the system to be hyperbolic, the decomposition of the system
into its eigenfields can be crucial in developing appropriate boundary
conditions. On a finite domain, boundaries have always been a serious problem
in numerical relativity.  But in a hyperbolic system one has detailed
knowledge of which quantities are propagating in which directions, and
also their speeds.  This allows a natural identification of radiative
variables.  This information is crucial in formulating conditions at
boundaries that allow outgoing quantities to actually escape from the
system (``outgoing radiation conditions''), while providing ways to
avoid generating unphysical and unwanted signals that propagate inward
from the boundaries (``no incoming radiation conditions'').  This is
useful not only at the outer boundary, but also it may be especially
important in the case of black holes where boundary conditions are
imposed on the horizon (``apparent horizon boundary conditions'', or
``AHBC'').  In AHBC, which exploits the causal properties of the
spacetime to chop out singular regions inside the black hole, a
detailed knowledge of the causal structure of the entire system of
equations is very important, and can be provided through a
hyperbolic treatment.

{\em (iii) Gauge modes.} Another important aspect of this way of
writing the system is the identification of gauge modes.  For slicings
that ensure hyperbolicity, not only are physical degrees of freedom
identified, which must propagate at the speed of light, but also
special gauge modes, which are related to the choice of slicing, are
naturally singled out.  This separation of physical from gauge effects
may play an important role in devising appropriate gauges or in
interpreting numerical results.

{\em (iv) Theoretical analysis.} Because hyperbolic systems of
conservation laws have been studied for many years, much is known
about their theoretical properties, the existence of solutions, the
treatment of boundaries, the conditions under which shocks can
develop, etc.~\cite{L&W,R&M,Godunov,Sod,Leveque}. With the
entire set of Einstein equations now cast in this form, for a variety
of practical gauge conditions, they can be studied with this body
knowledge in mind.

The starting point for this new formulation is the standard 3+1
decomposition of spacetime~\cite{Lich,Choquet,ADM}), which clearly
separates the dynamical degrees of freedom from the gauge ones: the
lapse function $\alpha$ and the shift components $\beta^i$.  In our
previous Letter~\cite{PRL95} we considered in detail only the zero
shift case just for simplicity.  We complete here the presentation by
considering the arbitrary shift case.  The term ``arbitrary'' here
means that one can prescribe the shift as a given spacetime function.
It does not mean that one can naively prescribe it by a local
relationship with dynamical quantities (like one does with the lapse
when choosing for instance harmonic slicing), because this would turn
the shift into a dynamical quantity and then the hyperbolicity of the
complete system would hold only for some specific choices.  We will
instead keep the shift as a purely kinematical degree of freedom which
can be adapted to every specific problem.  This is a subtle, but
important, point that will be discussed further below.

In order to cast the evolution system into first order hyperbolic form, we
introduce three extra dynamical quantities $V_i$ with evolution equations
provided by the momentum constraint. This approach is different from the
classical one, where harmonic coordinates are enough to get
hyperbolicity~\cite{Foures}. This is the price to pay for having an arbitrary
shift, because one can no longer use the three shift components to eliminate
unwanted terms in Einstein equations. The gauge-independent alternative of
using the three momentum constraints is more adapted to numerical applications.

This use of the momentum constraints is a shared feature in many new
hyperbolic formalisms~\cite{timederiv,Fritelli}.  Some of
these~\cite{timederiv} are variations on the idea from Y. Choquet and
T.  Ruggeri~\cite{Choquet83} of taking an extra time derivative to get
a third order evolution system which can then be written into first
order form.  Others prefer to take an extra space derivative of the
Einstein field equations~\cite{spacederiv} to take advantage of the
Bianchi identities.  In any case, the extra derivatives multiply the
number of independent quantities to be evolved.  Our approach,
instead, uses only three extra quantities and contains no extra
derivative of any kind.  In many senses, it is similar to the one
recently developed by S. Fritelli and O. Reula~\cite{Fritelli}.

Another shared feature in all the new hyperbolic formalisms is that
all of them allow harmonic slicing of the spacetime (harmonic time
coordinate).  This implies a local relationship between the lapse
function and the space volume element, so that the lapse becomes a
dynamic degree of freedom.  In our formalism, we generalize the
harmonic condition to a much wider set of dynamical slicings,
including the ones which have been actually used in successful
numerical applications~\cite{Bernstein,2DBH}.  To be more specific,
let us remember that in the zero shift case, harmonic slicing amounts
to a linear relationship between lapse and space volume element,
whereas our general case amounts to the lapse being any monotonically
increasing function of the space volume element.  An interesting point
is that this condition not only ensures hyperbolicity, but also
singularity avoidance.  We extend below to our general case a previous
proof for harmonic slicing~\cite{PRD88}.

In a recent work~\cite{Miguel}, the question of whether a dynamical
gauge condition can introduce the so-called ``coordinate shocks'' has
been raised.  Although our formalism~\cite{PRL95} has been used for
simplicity to illustrate the point, this problem is inherent to
harmonic slicing itself, and therefore to all formulations of the
equations, whether they follow the ADM approach, the new hyperbolic
formalisms~\cite{timederiv,Fritelli,spacederiv}, or any other
formulation.  The same problem has been detected even in numerical
codes based in non-hyperbolic systems but using dynamical
gauges~\cite{Laguna}.  More work is needed to understand the
implications of that effect, but in any case a way around this problem
is to use maximal slicing~\cite{York72}.  In that way one gets a
coupled elliptic-hyperbolic system in which the gauge degree of
freedom is elliptic so no coordinate shocks can appear due to the
slicing.  This could explain why codes based on maximal slicing are
usually more robust.

A final important point in our introduction is the choice of an
evolution system.  In our previous Letter~\cite{PRL95} we considered
only the standard choice, arising from the space components of the
Ricci tensor $R_{ab}$.  As we detail in section \ref{sec:evolution}
below, there is actually a one parameter family of physically
equivalent evolution systems which are all hyperbolic for the same
gauge choices with the same characteristic speeds: we provide the
eigenvectors for all of them.  Far from being a mathematical
curiosity, this choice freedom allows one to select the system in that
family which is free from Newtonian contributions.  By this we mean
that the evolution of the gravitational field is a purely relativistic
effect: in Newtonian gravity there is no evolution and the
gravitational field can be computed at every instant by integrating an
elliptic equation (the Poisson equation, which can be understood as
the Newtonian limit of the Hamiltonian constraint).  In all but one of
the evolution systems in that family the general relativistic
dynamical terms are mixed with pure Newtonian contributions.  The only
evolution system in which this does not happen turns out to be
different from the standard one: it arises from the space components
of the Einstein tensor $G_{ab}$ and then we will call it the
``Einstein system''.  This is crucial for numerical applications
either to weak field problems or at the outer boundary of a finite
difference grid.

\section{The Formalism}
\subsection{Space plus time decomposition}

In order to clarify the differences between the new evolution system
we propose and its predecessors, we first review the standard
evolution system.  The Einstein field equations consist of a
non-linear system of ten second order partial differential equations
when written in terms of the spacetime metric components $g_{\mu\nu}$.
In order to study the causal structure of this system, we will use a
time coordinate $t$ to label the evolution.  This amounts to
introducing a ``lapse'' function $\alpha$ relating $dt$ with the
proper time interval between the $t=constant$ hypersurfaces.

The following study depends on the actual choice of this time coordinate, so
that we will consider changes of the space coordinates only. In this sense, it
is clear that the lapse function $\alpha$ is a scalar quantity and the
3D ``induced metric'' $\gamma_{ij}$ on every constant $t$
hypersurface is a tensor quantity. It is then more convenient to write down
the line element in the following way
(3+1 decomposition~\cite{Lich,Choquet,ADM}):
\begin{equation}
ds^2 = -\alpha^2\;dt^2 + \gamma_{ij}\;(dx^i+\beta^i\;dt)\;(dx^j+\beta^j\;dt)
\;,	\label{metric}
\end{equation}
where the shift $\beta^i$ is related to the choice of space coordinates on
every $t=constant$ hypersurface.

Another important tensor quantity is the extrinsic curvature $K_{ij}$ (second
fundamental form) of the hypersurfaces, which can be expressed just as the
proper time derivative of the induced metric, taken along the normal lines:
\begin{equation}
(\partial_t - {\cal L}_\beta) \gamma_{ij} = -2\alpha\;K_{ij}  \;.
\label{Kij}
\end{equation}
Einstein's equations can then be expressed in terms of the set
of variables
\begin{equation}
(\alpha\;,\;\;\beta^i\;,\;\;\gamma_{ij}\;,\;\;K_{ij})     \;,
\label{3+1 vars}
\end{equation}
and it can be seen that no time derivative of the lapse nor the shift
appears into the resulting system. So far this is just the standard 3+1
approach~\cite{Lich,Choquet,ADM}.

We can either consider these kinematical quantities $\alpha$,
$\beta^i$ as being arbitrarily prescribed or we can provide
supplementary equations for them.  In this work we shall
take a combined approach, by supposing that the shift components
$\beta^i$ are known spacetime functions (we took them to be zero in
our previous Letter~\cite{PRL95} just for simplicity), whereas
we choose to evolve the lapse $\alpha$ according to
\begin{equation}
(\partial_t - \beta^k\partial_k) ln\,\alpha = -\alpha\;Q\; ,
\label{alphadot}       \\
\end{equation}
where the function $Q$ will be given later.  This form will turn out
to encompass many common choices of lapse.

\subsection{Evolution systems}
\label{sec:evolution}
The evolution of $K_{ij}$ is given by a set of six evolution
equations obtained from Einstein equations. For instance, the
space components of the four-dimensional Ricci tensor $^{(4)}R_{ij}$ can be
written~\cite{Choquet}
\begin{equation}
(\partial_t  - {\cal L}_\beta) K_{ij} = -\alpha_{i;j} + \alpha\;
[^{(3)}R_{ij} - 2K^2_{ij} + tr\,K\;K_{ij} - ^{(4)}R_{ij}] \; ,
\label{standard_evolution}
\end{equation}
where index contractions and covariant derivatives are with respect to
the induced metric $\gamma_{ij}$, and the three-dimensional Ricci
tensor constructed from the induced metric is denoted by $^{(3)}R_{ij}$.
This set of equations, together with (\ref{Kij}), is taken to be the
standard evolution system.  We shall call it in what follows the ``Ricci
evolution system'' for the set of variables~(\ref{3+1 vars}).

The remaining four Einstein equations are constraints, which can be
easily identified: the ``energy'', or Hamiltonian constraint
\begin{equation}
2\alpha^2 \; G^{00} = {}^{(3)}R + (tr\,K)^2 - tr(K^2) \;,
\label{energy_constraint}
\end{equation}
where $^{(3)}R$ is the trace of the three-dimensional Ricci tensor,
and the ``momentum constraint''
\begin{equation}
\alpha \; G^0_{\;i} = K^k_{\;i;k} - \partial_i(tr\,K) \;.
\label{momentum_constraint}
\end{equation}
These constraint equations are first integrals of the evolution
system.  They are then redundant provided that they are imposed on the
initial data (otherwise one would get unphysical solutions).

It is not often appreciated that, although it is the standard
evolution system, the Ricci system (\ref{standard_evolution}) is not
convenient for many numerical applications.  One way of seeing this is
to look at the space components of the Ricci tensor for a perfect
fluid,
\begin{equation}
^{(4)}R_{ij} = 8\pi [(\mu + p) u_iu_j + 1/2\,(\mu -p)\,\gamma_{ij}] \;\; .
\end{equation}
where $\mu$ is the total energy density of the fluid, $p$ is the
pressure, and $u_{i}$ is its fluid 3-velocity.  Notice that the second
term contains a contribution from the energy of the fluid which does
not vanish in the Newtonian limit, where the 3-velocities are taken to
be small compared to one and the pressure small compared to the energy
density.

This means that the system (\ref{standard_evolution}) has a Newtonian
contribution from the energy density which, allowing for the energy constraint
(\ref{energy_constraint}), is to be compensated with other Newtonian
contributions in the geometry terms. This situation
can be very inconvenient either in ``post-Newtonian'' (moderately
relativistic) scenarios, where the small relativistic evolution
effects can easily be masked by the truncation errors of the larger
newtonian terms, or even in strong field scenarios where it can be
difficult to preserve the constraints at the boundaries.

To remedy this problem, let us note that an evolution equation plus a
constraint leads to another evolution equation.  Using the relation
\begin{equation}
^{(4)}R_{ij} = G_{ij} - 1/2\,(-\alpha^2G^{00} + tr\,G)\,\gamma_{ij} \;\; ,
\end{equation}
where we have noted $tr\,G = \gamma^{ij}G_{ij}$, we will combine the energy
constraint (\ref{energy_constraint}) with (\ref{standard_evolution}) to
cancel out the $G^{00}$ term, obtaining a different evolution system
\begin{eqnarray}
(\partial_t - {\cal L}_\beta) K_{ij} = -\alpha_{i;j} + \alpha\;
[{}^{(3)}R_{ij} -2K^2_{ij} + tr\,K\;K_{ij} - G_{ij}]&\;&
\nonumber \\
-\alpha/4\;\gamma_{ij}\;[{}^{(3)}R - tr(K^2) + (tr\,K)^2 - 2tr\,G]&\;&,
\label{post-newtonian}
\end{eqnarray}
which is equivalent to the one arising from the space components $G_{ij}$ of
the Einstein tensor.

We shall call it in what follows the ``Einstein evolution
system''. The matter terms in the perfect fluid case can be computed now from
\begin{equation}
G_{ij} = 8\pi[(\mu+p) u_iu_j + p\,\gamma_{ij}] \;\; ,
\end{equation}
so that they vanish in the Newtonian limit. This
``Einstein system'' has been found to be useful in tests of
hydrodynamic evolution\cite{ValenciaWork}, but it is important not only in
the matter case.  The use of the Einstein system
(\ref{post-newtonian}) turns out to be very important to obtain the long term
evolution for a {\em vacuum} 1D black hole that we presented in our
previous Letter, and will be discussed in detail in a future paper in
this series.

The two systems (\ref{standard_evolution}) and (\ref{post-newtonian})
are not equivalent: they can, in principle, have different solutions.
However, the physical solutions (the ones verifying the constraints)
are common to both systems.  Physics is not affected, of course, but
the mathematical structure can be modified by the choice of the
evolution system among the infinitely many combinations of the Ricci
system with the energy and/or momentum constraints.  As we have
suggested that the use of the energy constraint is important for
accuracy, we will see below that the use of the momentum constraint is
crucial to ensure hyperbolicity ~\cite{Choquet83,PRL92}.  This
provides {\em a posteriori} a good criterion for choosing a particular
evolution system among the many possibilities.

\subsection{A first order evolution system}
The evolution systems (\ref{standard_evolution},
\ref{post-newtonian}) are first order in time, but second order in space.
To obtain systems which are also of first order in space, we will follow the
standard procedure by introducing auxiliary
variables which correspond to the space derivatives,
\begin{equation}
A_k = \partial_k ln\,\alpha\, ,
\;\;\;\; B_k^{\;i} = 1/2\;\partial_k \beta^{i}
\;\;\;\; D_{kij} = 1/2\;\partial_k \gamma_{ij}\;.
\label{Ds}
\end{equation}
Note that the shift components are given at
every instant so that the space derivatives $B_i^{\;j}$ are known.
The evolution equations for the remaining quantities can be obtained by taking
the time derivative of (\ref{Ds}) and interchanging the order of space and time
derivatives:
\begin{eqnarray}
\partial_t  A_k &+& \partial_k [-\beta^r A_r + \alpha\;Q] = 0
\label{evolution_A}       \\
\partial_t  D_{kij} &+& \partial_k [-\beta^r D_{rij}
	+ \alpha\;(K_{ij}-s_{ij})] =  0 \;,
\label{evolution_D}
\end{eqnarray}
where we have used the shorthand
\begin{equation}
s_{ij} = (B_{ij}+B_{ji})/\alpha,
\end{equation}
and for notational convenience, we have also written $B_{ij} =
\gamma_{ik}B^{\;k}_{j}$, even though $B_{ij}$ is not a tensor quantity.

Note that we have used here the ordering freedom of space derivatives
in a different way than we did in our previous Letter~\cite{PRL95}, where
Eq.~(\ref{evolution_D}) was written as
\begin{equation}
\partial_t  D_{kij} + \partial_r [-\beta^r D_{kij}
	+ \alpha\;\delta^r_k\;(K_{ij}-s_{ij})] =
	(2B_k^{\;r}-\alpha\;tr\,s \;\delta^r_k)\;D_{rij}\;,
\end{equation}

The present choice (\ref{evolution_D}) is more suitable for numerical
applications when the shift does not vanish, as it does not
introduce extra sources. The same criterion leads us to write down
(\ref{Kij},\ref{alphadot}) in the following way
\begin{eqnarray}
\partial_t \gamma_{ij} &=& -2\alpha\;(K_{ij}-s_{ij}) + 2\beta^r\,D_{rij} \;,
\nonumber \\
\partial_t ln\,\alpha &=& -\alpha\;Q + \beta^r\,A_r\; .
\label{background}
\end{eqnarray}

So far, Eqs.(\ref{evolution_A}-\ref{background}) have been written in a
first order balance law form
\begin{equation}
\partial_t {\bf u} + \partial_k F^k_-{\bf u} = S_-{\bf u}
\label{3Dbalance}
\end{equation}
where the vector ${\bf u}$ displays the set of variables and both
``fluxes'' $F^k$ and ``sources'' $S$ are vector valued functions.
Our goal will be write the entire system of evolution equations in
this form.  We introduce the additional quatities
\begin{equation}
V_i = D_{ir}^{\;\;r}-D^r_{ri}\;,
\label{vector}
\end{equation}
where again even though the $D_{ijk}$ are not components of a tensor,
we raise indices in the usual way with the three-metric $\gamma_{ij}$.
Then after extensive manipulation the evolution equations for the
extrinsic curvature components (\ref{standard_evolution}) can also be
put in the first order balance law form given by Eq.~(\ref{3Dbalance}).

This almost completes the system, for which the nonzero fluxes are
\begin{eqnarray}
\label{fluxes}
F^k_-A_k &=& -\beta^r A_r + \alpha\;Q  \\
F^k_-D_{kij} &=& -\beta^r D_{rij} + \alpha\;(K_{ij}-s_{ij}) \\
F^k_-K_{ij} &=& -\beta^k\,K_{ij} + \alpha [D^k_{ij} - n/2\;V^k\;\gamma_{ij}
\\ &\;& + 1/2\;\delta^k_i\;(A_j+2\,V_j-D_{jr}^{\;\;r})
\nonumber \\ &\;& + 1/2\;\delta^k_j\;(A_i+2\,V_i-D_{ir}^{\;\;r})]  \;\;,
\nonumber
\end{eqnarray}
where the free parameter $n$ allows one to select a specific evolution system
(it is zero for the Ricci system and one for the Einstein system).
The nonzero source terms are those appearing in (\ref{background}) and
\begin{eqnarray}
S_-K_{ij} &=& 2(K_{ir}B_j^{\;r}+K_{jr}B_i^{\;r}-K_{ij}B_r^{\;r}) \nonumber \\
   &&+\alpha \{ -^{(4)}R_{ij}- 2K_i^{\;k}K_{kj}+tr\,K\;K_{ij} \nonumber  \\
       &&-\Gamma^k_{\;ri}\Gamma^r_{\;kj}+2D_{ik}^{\;\;r}D_{rj}^{\;\;k}
    +2D_{jk}^{\;\;r}D_{ri}^{\;\;k} +\Gamma^k_{\;kr}\Gamma^r_{\;ij}\nonumber \\
          &&-(2D_{kr}^{\;\;k}-A_r)(D_{ij}^{\;\;r}+D_{ji}^{\;\;r})
        +A_i(V_j-1/2\;D_{jk}^{\;\;k})      \\
       &&+A_j(V_i-1/2\;D_{ik}^{\;\;k})
	-nV^kD_{kij}   \nonumber \\
       &&+n/4\;\gamma_{ij}[-D_k^{\;rs}\Gamma^k_{\;rs}
        +D_{kr}^{\;\;r}D^{ks}_{\;\;s}-2\;V^kA_k+tr(K^2)-(tr\,K)^2
	+2\alpha^2\;G^{00}] \} \;\;.
	\nonumber
\end{eqnarray}

So far we have achieved a balance law formulation which will be valid
for any gauge choice (lapse and shift).  Already this is something
useful, as many numerical methods, such as the genuine
MacCormack scheme~\cite{maccormack}, have been devised explicitly for
such a system.  In particular we note that without this balance law
formulation, one {\em cannot} apply a method like MacCormack, which
was designed to treat not only the time evolution, but also the
fluxes and sources, in a specific way.  Previous applications of a
``MacCormack-like'' method, as in Ref.~\cite{Bernstein89}, used only
the time evolution part of this scheme, leaving the most important
spatial part of the system to be treated without regard for the
particular structure of the equations.

However, as important as the balance law formulation is, additional
benefit could be gained if the system would be actually hyperbolic.
This means that the entire system of balance law equations
(\ref{3Dbalance}) can be diagonalized, with a complete set of
eigenvectors with real eigenvalues.  This is not yet the case, mainly
because of the combinations (\ref{vector}) arising in the flux terms
(\ref{fluxes}).  They could be eliminated by a suitable shift choice
(imposing harmonic space coordinates, for instance), but we prefer to
deal with an arbitrary shift and we will proceed in a different way.

The three combinations defined by (\ref{vector}) are very interesting
quantities.  One can actually compute their time derivative from
(\ref{evolution_D}) and make use of the momentum constraint
(\ref{momentum_constraint}) to obtain for these three combinations
evolution equations of the balance law form (\ref{3Dbalance}) with
\begin{eqnarray}
\label{evolution_V}
F^k_-V_i &=& -\beta^k V_i + B^k_{\;i} - B_i^{\;k} \nonumber \\
S_-V_i &=& \alpha\;[\alpha\;G^0_{\;i}
			+ A_r\;(K^r_{\;i}-tr\,K\;\delta^r_i)  \\
                      &&+ K^r_{\;s}(D_{ir}^{\;\;s}-2D_{ri}^{\;\;s})
                       - K^r_{\;i}(D_{rs}^{\;\;s}-2D_{sr}^{\;\;s})]
	\nonumber\\
	&&+ 2(B_i^{\;r} - \delta_i^r\;tr\,B)\;V_r
          + 2(D_{ri}^{\;\;s}-\delta^s_i\;D^j_{\;jr})B^r_{\;s}.
	\nonumber
\end{eqnarray}
Then one can relax the algebraic condition (\ref{vector}) and consider
$V_i$ as a set of supplementary independent quantities to be evolved
according to their evolution equations (\ref{evolution_V}).  The
vector array ${\bf u}$ representing the independent quantities
satisfying the balance law equations (\ref{3Dbalance}) will then contain
the following $37$ functions:
\begin{equation}
{\bf u} = (\alpha\;,\;\;\gamma_{ij}\;,
\;\;K_{ij}\;, \;\; A_i\;,\;\;D_{rij}\;,\;\;V_i)     \;,
\label{full vars}
\end{equation}
so that the condition (\ref{vector}) can be now considered as an
algebraic constraint which will hold if and only if the momentum constraint is
satisfied. This is the key point to get a hyperbolic system, and it has nothing
to do with the coordinate gauge: it is just making a free use of the momentum
constraint, a feature which is shared by other recent hyperbolic
formulations~\cite{timederiv,Fritelli}.  The conditions under which
this system is actually hyperbolic will be given below, along with the
explicit eigenvectors and eigenvalues (characteristic speeds) and the
diagonalized system.

\subsection{Invariant algebraic slicing}

Before we can complete our analysis, we need to
know how the ``slicing source function'' $Q$, and how the shift
vector $\beta^{i}$, depend on the fields to
be evolved.  We will use the lapse function degree of freedom to
specify a time coordinate, which amounts to specifying $Q$.  This will
be done by relating the lapse to the space metric coefficients
(dynamical lapse), but keeping the freedom of choosing arbitrary space
coordinates on every slice (kinematical shift).  We will demand then
our lapse to be an algebraic condition, invariant under any
transformation of the space coordinates on every slice.  We must use
then scalars, like $\alpha$, $Q$, $tr\,K$ and their proper time
derivatives.  If we restrict ourselves to scalars containing no
derivatives of the metric coefficients, we can play only with $\alpha$
and we get either a ``geodesic slicing'' ($\alpha = constant$) or one
of its generalizations.  This is too restrictive, as we will see
later.  If we allow also for first order derivatives of the metric, we
have also $Q$ and $tr\,K$ at our disposal.  As we have seen in the
previous section, the principal part of the evolution system is
quasilinear, so let us take a generic quasilinear homogeneous
condition~\cite{PRL95}
\begin{equation}
Q = f(\alpha)\;tr\,K
\label{linear}
\end{equation}
where $f$ is an arbitrary function.

The geodesic slicing is then included as a subcase with $f=0$.
The $f=1$ case corresponds to the ``harmonic
slicing''~\cite{Choquet83,PRD88} (the resulting time coordinate is
harmonic). Another interesting case is the ``1+log''
slicing~\cite{Bernstein,2DBH}, obtained when $f=1/\alpha$; it mimics maximal
slicing near a singularity, when the lapse collapses to zero. (Here we
have considered the case $\alpha \propto 1+log(\sqrt{\gamma})$, which differs
slightly from that considered in~\cite{Bernstein,2DBH}.  Both cases
are included in this class of slicings.)

The slicing condition (\ref{linear}) can be integrated in normal coordinates
(zero shift) to obtain, up to some integration constant,
\begin{equation}
\sqrt{\gamma} = F(\alpha)\;,
\label{inteq}
\end{equation}
where $F$ is an arbitrary function.  This shows the
generality of this condition, which is somehow hidden in its invariant
form (\ref{linear}).

The widely used maximal slicing~\cite{Lich,York72} ($tr\,K = 0$) is
included also as a limiting case when $f$ diverges ($F$ is constant).
It is a very special case because the lapse $\alpha$ is given by an
elliptic condition, so that the evolution system becomes a coupled
hyperbolic-elliptic system.  Moreover, $\alpha$ is no longer related
to the space volume element $\sqrt{\gamma}$ (which is actually
constant in the zero shift case) and the recent discussion of
``coordinate shocks''~\cite{Miguel} does not apply to maximal slicing.

It is worth studying the behavior of the slicings defined by (\ref{linear})
when the space volume element $\sqrt{\gamma}$ goes to zero. This will be a
singularity of the slicing, but it is more convenient to view it in an
equivalent way, as a singularity of the congruence of time lines normal to
the slicing, so that we can use the integral form (\ref{inteq}). Let us
suppose that this singularity occurs after a finite proper time interval
$\tau_s$ away from our initial time slice. The elapsed coordinate time
will be given by the integral
\begin{equation}
\Delta t = \int_0^{\tau_s} \frac{d\tau}{\alpha} \;,
\label{integral}
\end{equation}
so that an obvious necessary condition for singularity avoidance is that the
lapse function vanishes before or at the singularity (lapse collapse), because
otherwise the integral (\ref{integral}) will be finite and this means that the
singularity will be reached in a finite coordinate time.

If the lapse vanishes (lapse collapse) at $\tau_0 < \tau_s$, the
slicing is said to have a ``limit surface'' and it stops before reaching the
singularity: this happens for instance with maximal~\cite{Estabrook73}
or ``1+log'' slicing.  If the lapse vanishes precisely at $\tau_0 = \tau_s$,
singularity avoidance would mean that the improper integral (\ref{integral})
diverges (one does not reach $\tau_s$ in a finite time). One can obtain a
sufficient condition for singularity avoidance for ``focusing
singularities''~\cite{PRD88}, that is
when the space volume element vanishes at a bounded rate, so that
\begin{equation}
	  |\partial_\tau \sqrt{\gamma}|_{\tau_s} =
          |F'(\alpha)\;\partial_\tau \alpha|_{\tau_s} < B \;\; ,
\end{equation}
and it is clear that if we assume strict monotonicity of $F$ at the
singular point
\begin{equation}
          |F'(\alpha)|_{\tau_s} \neq 0 \;\; ,
\end{equation}
that would imply that the lapse itself vanishes at a bounded rate
\begin{equation}
          |\partial_\tau \alpha|_{\tau_s} < B' \;\; ,
\end{equation}
and the improper integral (\ref{integral}) would not converge: the
singularity can not be reached in a finite coordinate time. It follows
that focusing singularities are avoided by strictly monotonic choices of $F$,
like the ones that ensure hyperbolicity, as we will see below.

\subsection{Causal structure of the evolution system}
In what follows we will analyze the causal structure of the set of
equations we have derived. It will turn out that under certain
conditions, the system is hyperbolic, allowing a better understanding
of the theoretical properties of the system that also permits yet more
powerful numerical methods to be applied to the Einstein equations.
In this section we consider the shift vector $\beta^{i}$ as a known
function of spacetime.  In the next section we discuss the shift and
its effect on the system more fully.

The causal structure of a first order system is given by its
principal, or transport, part.
The source terms contain no space derivatives, so
that the principal part is given by the flux terms
\begin{equation}
\partial_t {\bf u} + \partial_k F^k_-{\bf u} = 0 \;\; .
\label{3Dtransport}
\end{equation}
In this kind of analysis it is essential to write the entire system so
that the source terms contain no derivatives of the fields.
Otherwise, by manipulating the flux (derivative) and source terms one
could apparently change the causal structure of the system at will.
We can consider the transport part separately by splitting the
evolution described by (\ref{3Dbalance}) into two separate processes:
the first one is the transport process described by
(\ref{3Dtransport}), and the second one is the sources contribution,
given by the following system of {\em ordinary} differential equations
\begin{equation}
\partial_t {\bf u} = S_-{\bf u} \;\; .
\label{3Dsources}
\end{equation}

This conceptual splitting can be easily implemented in numerical applications.
If we note by $E(\Delta t)$ the numerical evolution operator for system
(\ref{3Dbalance}) in a single timestep, we get that, up to second order
accuracy in $\Delta t$,
\begin{equation}
E(\Delta t) = S(\Delta t/2)\;T(\Delta t)\;S(\Delta t/2)
\label{splitting}
\end{equation}
where $T$, $S$ are the numerical evolution operators for systems
(\ref{3Dtransport}) and (\ref{3Dsources}), respectively.  This is
known as ``Strang splitting''~\cite{recipes}.

Note that, according to (\ref{background}) the evolution equations for
the lapse and the induced metric have no flux terms.  This means that
we can regard the transport step as the propagation of a reduced set
of $30$ quantities
\begin{equation}
{\bf u} = (K_{ij}\;, \;\; A_i\;,\;\;D_{rij}\;,\;\;V_i)
\label{reduced vars}
\end{equation}
in an inhomogeneous ``background''.  The equation (\ref{3Dtransport})
is linear in the quantities (\ref{reduced vars}), and this is a key
point in what follows.  This means that {\em during the transport
step} the ``background'' quantities, which evolve according to
Eq.~(\ref{3Dsources}), are fixed.

The standard procedure for studying the causal structure of first order
systems starts by choosing a fixed space direction. Only space
derivatives along this direction will be considered, so that the resulting
system is actually one-dimensional. This procedure does not
match the usual one for second order equations, where there is no need for
choosing {\em a priori} a direction and all derivatives are dealt with
simultaneously. The first order formalism, in contrast, allows one to treat
one direction at a time, and this ``locally one-dimensional'' (LOD) approch
is useful both for theoretical analysis and numerical applications.

Let us begin our LOD analysis by taking for instance our space direction along
the $x^k$ coordinate axis. We will then neglect all fluxes along the other
directions $x^{k'}$. It follows that, apart from the background metric
coefficients, the $14$ quantities
\begin{equation}
A_{k'}\;,\;\;D_{k'ij}\;,\;\;
\;\; (i,\,j=1\,,2\,,3\;\;k'\neq k)
\end{equation}
have no flux along the $x^k$ direction, so that they are
characteristic fields propagating along the time lines (zero
characteristic speed).  Propagation along time lines is much more
convenient for numerical applications than propagation along normal
lines (speed $-\beta^k$) as we had in our previous Letter
~\cite{PRL95}.  (As we will see below, propagation along time lines
can be treated by methods for ODE's.) They only coincide in the zero
shift case.

Rather than studying the transport properties of the ${\bf u}$'s, the
evolution of the remaining $16$ quantities is more easily studied by
taking their {\em fluxes} (\ref{fluxes}) to be the basic quantities.
Hence we rewrite the LOD transport equations using the fluxes along the
selected direction $x^k$ as the basic quantities.
We find that (the principal part of) the resulting equations
can be written as a system of 16 one-dimensional advection equations
(no sum in $k$)
\begin{equation}
\partial_t \left( 
\begin{array}{c} 
F^k_-V^k       \\
F^k_-V_{k'}     \\
F^k_-K_{ik'}    \\
F^k_-D_{kik'}   \\
F^k_-K_r^{\;r}  \\
F^k_-A_k        \\
F^k_-D_{kr}^{\;\;r}
\end{array} \right) + {\bf A}\; \partial_k \left(
\begin{array}{c}
F^k_-V^k       \\
F^k_-V_{k'}     \\
F^k_-K_{ik'}    \\
F^k_-D_{kik'}   \\
F^k_-K_r^{\;r}  \\
F^k_-A_k        \\
F^k_-D_{kr}^{\;\;r}
\end{array} \right) = 0 \; ,
\label{advection}
\end{equation}
where ${\bf A}$ is the characteristic matrix of this reduced system
\begin{equation}
{\bf A} = \left( 
  \begin{array}{ccccccc}
  -\beta^k & 0 & 0 & 0 & 0  & 0 & 0 \\
  0 & -\beta^k & 0 & 0 & 0  & 0 & 0 \\
  -n/2\,\alpha\,\gamma_{ik'}
    & \alpha\,\delta^k_i & -\beta^k & \alpha\,\gamma^{kk} & 0  & 0 & 0 \\
  0 & 0 & \alpha & -\beta^k & 0  & 0 & 0 \\
  (2-3n/2)\,\alpha & 0 & 0 & 0 & -\beta^k & \alpha\,\gamma^{kk} & 0   \\
  0 & 0 & 0 & 0 & \alpha\,f & -\beta^k & 0  \\
  0 & 0 & 0 & 0 &\alpha & 0 & -\beta^k
\end{array} \right)  \;\; .\label{matrix}
\end{equation}
Its eigenvalues are then the ``characteristic speeds''. The corresponding
right eigenvectors are the ``characteristic fields''.

Let us list the $16$ characteristic fields associated with
(\ref{matrix}):
\begin{itemize}
\item The three quantities $F^k_-V_i$ plus the single quantity
$F^k_-A_k-f F^k_-D_{kr}^{\;\;r}$, which propagate along normal lines
(speed $-\beta^k$).
\item The ten combinations
	\begin{equation}
	F^k_-K_{ik'} \pm \sqrt{\gamma^{kk}}\,
		[F^k_-D_{kik'}+ (\delta^k_i\,F^k_-V_{k'}
				-n/2\,\gamma_{ik'}F^k_-V^k)/\gamma^{kk}] \;\;,
        \label{light}
	\end{equation}
        which propagate along light cones (speed $-\beta^k \pm
	\sqrt{\gamma^{kk}}$, respectively).
\item The two combinations
	\begin{equation}
	\sqrt{f}\,F^k_-K_r^{\;r} \pm \sqrt{\gamma^{kk}}\;[F^k_-A_k
					+(2-3n/2)\,F^k_-V^k/\gamma^{kk}]
	\label{gauge}
	\end{equation}
        which propagate, respectively,  with the gauge dependent speed
	$-\beta^k \pm \sqrt{f\;\gamma^{kk}}$ (``gauge speed'').
\end{itemize}

A system is said to be hyperbolic if all the characteristic speeds are
real and the characteristic matrix can be fully diagonalized (see for
instance Ref.~\cite{Leveque}).  This is our case provided that $f>0$
(note that if $f=0$, as in the geodesic case, the last combination
(\ref{gauge}) contains only one independent quantity and the set of
eigenfields is no longer complete).  Gauge speed coincides with light
speed only in the harmonic case ($f=1$).  It becomes infinite for a
maximal slicing, which can be considered as a limiting case of our
condition (\ref{linear}).

The advantage of having a hyperbolic system is that we know now
explicitly which combination is propagating forward or backward along
the selected direction. Suppose for instance that one is using a 3D
cartesian finite difference grid with vanishing shift at the outer
boundaries. It follows that the five combinations one gets from
(\ref{light}) by using the plus (respectively minus) sign are entering
the grid through the left (respectively right) outer boundary along
the $x^k$ direction. The same thing happens with the gauge combination
(\ref{gauge}). This information should be very valuable when devising
boundary conditions, as we will show in a future publication.

The very existence of ``gauge speeds'' is a remarkable result.  One is
used to thinking that light cones are enough to determine the causal
structure of spacetime.  This is true if we refer only to the
invariant features.  But the evolution system evolves spacetime
together with the coordinate system we are using to label it (the
dynamical lapse in our case).  For instance, maximal slicing is
associated with an infinite gauge speed (as it must be, because both
the lapse and its derivatives are provided by an elliptic equation).
These considerations single out the harmonic slicing, in which gauge
cones and light cones do coincide.  This further degeneracy simplifies
the causal structure of spacetime, but it is just accidental and we
see no reason to overlook the richer structure that arises in the
general case.

\subsection{The role of the shift}
\label{sec:shift}

The balance law evolution equations for the entire set of variables
(\ref{full vars}) is valid for any choice of lapse function $\alpha$
and shift vector $\beta^{i}$ whatever.  In the analysis of the causal
structure, we considered very carefully the effect of the choice of
lapse, and showed that for a large family of conditions, the system is
actually hyperbolic.  But so far we have said little about the shift,
treating it as a given spacetime function.  In such a case it has no
dynamics and hence it does not enter in to the discussion of the
causal structure of the system, except through its connection with the
eigenspeeds of the other variables.

However, in practice in numerical relativity, the shift is {\em not}
prescribed ahead of time as a known function of the spacetime
coordinates.  Instead, one wants it to respond to the dynamics of the system,
and it is usually related in some explicit functional way to the metric
and extrinsic curvature variables themselves.  If the shift is taken
to be some explicit function of the other fields, and introduced to
the system in a way that it changes continuously as a dynamical
variable, this could affect the causal structure of the system.  In
this case, one would need to substitute this prescription into the
complete system and analyze the causal structure on a case by case
basis.  In all cases, the equations are still valid, but as far as the
causal structure is concerned, for some cases the eigenfields and
eigenspeeds could change, or the hyperbolicity itself could be broken.
This statement should be true for any formulation of the Einstein
equations.  A careful analysis of many such cases, where the shift is
considered as a true dynamical variable, has been carried
out~\cite{Bona94c,Suen95b}.

On the other hand, the shift need not be considered as a dynamical
variable of the system.  One is free to choose it any time as one
likes.  For example, on a given time slice one could prescribe it to
be any arbitrary function of the other fields in the system, and hold
this fixed as long as one likes.  While the other variables in the
system evolve, the shift is held fixed in time.  Then the shift has no
dynamics, and cannot affect the causal structure of the system.  On a
later time slice, one may again choose the shift freely, and then
consider it to be fixed for the next period of evolution.  In fact,
one can do this as often as one likes, say on every time slice.
However, it is crucial to point out that this is {\em not} the same as
having a dynamic shift that can affect the causal structure of the
system.  In the latter dynamic case, the shift changes continuously as
the other fields evolve.  In the case we are considering, which we
could call a momentarily ``frozen shift'', the evolution of the fields
is {\em decoupled} from the development of the shift, and vice versa.
During the evolution of the fields the shift has no dynamics, and is a
known function of space.  This is a subtle, yet crucial point.  It is
not merely a point of view, but a key practical point to be made.  The
shift can be chosen in this way on every time step, but in the process
of evolving the fields from one time step to the next it must be
regarded as a fixed function. Note that we can not say the same about the
lapse, because this will amount to drop out the lapse derivatives $A_k$ (their
fluxes in Eq.~(\ref{advection})) from the list of dynamical quantities. This
means that we should then suppress the corresponding row and column in the
characteristic matrix~(\ref{matrix}), that will no longer be diagonalizable.

As in this treatment the shift is a known function of space for all
evolution steps, the previous analysis of the causal structure and
hyperbolicity carries through for {\em all} shift choices one cares to
make on all time steps.  However, this treatment has certain numerical
consequences that will be considered in the next section.

\section{Numerical methods}

In this section we describe in broad terms the kind of numerical
methods that can applied to this formulation of the Einstein
equations.  We will defer a detailed treatment, complete with
numerical examples, to future publications.  In the previous sections,
we have reformulated the evolution equations as a first order system
of balance laws, valid for any choice of lapse and shift, without
taking any additional derivative.  This allow one to use standard
numerical methods from Computational Fluid Dynamics: Lax-Wendroff,
staggered leapfrog, MacCormack~\cite{recipes}.  We have also seen
that for some choices of the lapse function the evolution system is
hyperbolic, so that we can use more advanced numerical
methods~\cite{Leveque} and have better control at the boundaries, as
we will see below.

To obtain a finite difference version of our equations, one can use the
numerical splitting approach as we outlined above
(See, e.g., Eq.~(\ref{splitting})). This has many advantages,
which we summarize here:
\begin{itemize}

\item The nonlinear terms in Einstein equations (the ones containing
products of first derivatives of the metric) appear only in the
sources step (\ref{3Dsources}), in the form of a coupled system of
ordinary differential equations.  In numerical tests we have
discretized this part using standard predictor-corrector
or modified midpoint methods ~\cite{recipes} to second order accuracy,
although many prescriptions are possible.  Note that this gives one
the possibility of using methods for stiff ODE's if the source terms
behave in this way.

\item The remaining terms (the principal part) are in Flux Conservative form
(\ref{3Dtransport}). This allows us to apply the high resolution methods which
have been developed for Computational Fluid Dynamics. We have discretized that
part using a second order TVD (Total Variation Diminishing) method
(\cite{Leveque}), although again many possibilities could be examined.

\item We can taylor the shift to fit our needs by choosing any profile
just after every sources step, but keeping it constant during the
whole transport step (``frozen'' shift).  This allows of course an
explicit prescription of the shift (we have used for instance a
parabolic shift profile to track the horizon of 1D black holes).  But
this also allows an indirect prescription, as it can be done for
instance via the solution of some elliptic equation (minimal
distortion shift, or similar conditions), exactly in the same way we
do when imposing the maximal slicing condition for the lapse.  The key
point is that the shift must be kept fixed during the transport step,
as described above, or else the analysis of the eigenfields would have
to be redone.

\item Allowing for (\ref{background}) and the previous considerations,
the metric coefficients do not evolve in
the transport step. This means that we are just evolving now the reduced set of
dynamical variables (\ref{reduced vars}) and the principal part
(\ref{3Dtransport}) is  linear in these quantities:
\begin{equation}
   F^k_-{\bf u} = {\bf A}^k (x)\;{\bf u} \;\;. \label{jaclin}
\end{equation}
We can then describe the
transport step by writing (\ref{3Dtransport}) in the form
\begin{equation}
\partial_t {\bf u} + \partial_k [{\bf A}^k (x)\;{\bf u}] = 0 \;\; ,
\label{reduced transport}
\end{equation}
because the metric coefficients contained in the matrix ${\bf A}^k$ are time
independent in the transport step. The resulting equation
(\ref{reduced transport}) can be then interpreted as a transport equation in an
inhomogeneous (but fixed) background. The alternative form
(\ref{advection}) for the evolution of the fluxes (\ref{jaclin})
follows then easily.

\end{itemize}

The complete numerical details of 1D and 3D codes developed to solve these
equations with a variety of advanced methods will be given
elsewhere.  It is our purpose to outline all details
necessary to understand both theoretical and practical aspects of our
approach, and to provide some simple examples of its power.

Before moving to more specific application in the 1D case, we further
discuss an important numerical aspect of the treatment of the shift.
As we have stressed, the equations we have presented are valid for
{\em any} choice of lapse and shift, and a host of numerical methods
may be applied as described.  For applications where hyperbolicity is
important, it is clear that the system is hyperbolic for a certain
family of slicings and a given spacetime function of shift.
Furthermore, any prescription for the shift, even if it is explicitly
a function of the fields ${\bf u}$ and their derivatives, can preserve
hyperbolicity if the system is treated appropriately by holding the
shift fixed during the transport step.

The numerical implication of this treatment is important.  It means
that if hyperbolicity is desired, in some cases the shift may only be
enforced numerically in a first order way during the evolution.  For
example, in a method that requires a prediction and correction of the
fields during the transport step, the shift must remain decoupled
(fixed) during both steps.  Hence although the fields may be evolved
forward in time in a second order way, in our present scheme the shift
itself is not necessarily implemented to second order accuracy in time.  This
issue {\em only} arises if one exploits information about the
eigenfields in the actual numerical implementation.  For example, in a
MacCormack method, which does not make use of hyperbolicity, this is
not an issue, nor is it an issue if the shift is a given spacetime
function. The convergence properties for complicated examples,
where the shift is prescribed as an explicit function of the fields
(say, minimal distortion shift), and where knowledge of hyperbolicity is
used in the numerical methods, will have to be investigated carefully
in the future.

\subsection{The one-dimensional case}
Let us further illustrate the advantages of our formulation by
outlining numerical methods in the one-dimensional case.  The 3D case
can be handled similarly by directional splitting.  To label the
values, we will use a superindex for the time level and a subindex for
the grid position.  For the transport equations,
\begin{equation}
\partial_t {\bf u} + \partial_x F_-{\bf u} = 0 \;\; .
\label{1D transport}
\end{equation}
we can then use flux-conservative numerical methods of the form
\begin{equation}
u^{n+1}_i = u^n_i
-{\Delta t \over \Delta x} (F^{n+1/2}_{i+1/2} - F^{n+1/2}_{i-1/2}).
\label{steptwo}
\end{equation}
These are two-step algorithms.  In the first step, one must compute
the fluxes at the ``interface points'' $x_{i \pm 1/2}$ at the
intermediate time level $t^{n+1/2}$ to a given accuracy (we will use
first order).  The final step (\ref{steptwo}) improves the overall
accuracy by one order (we get then second order overall accuracy).
The timestep is limited by the causality condition (also known as
Courant-Friedrichs-Lewy condition~\cite{recipes}), which states that
the maximal speed in the numerical algorithm is just ${\Delta x /
\Delta t}$ (one grid point at a time), providing an upper bound to the
time step $\Delta t$ to ensure that this speed is greater than the
characteristic speeds.

In order to compute the interface fluxes $F^{n+1/2}_{i+1/2}$, we will take
advantage of the fact that, allowing for (\ref{reduced transport}), the
transport equation (\ref{1D transport}) can be written as
\begin{equation}
\partial_t {\bf F} + {\bf A} (x)\;\partial_x {\bf F} = 0 \;\; ,
\label{1D advection}
\end{equation}
which is the one-dimensional version of the system of advection
equations (\ref{advection}).  This means that the eigenfields of the
characteristic matrix ${\bf A}$ (\ref{matrix}) propagate along
characteristic lines in either the forward or backward direction,
depending on the sign of their characteristic speed. The fluxes
can be first evaluated at every grid point, and then the
diagonal combinations (i.e., the eigenfields) are computed.  These diagonal
combinations are propagated to the grid interfaces, using (\ref{1D
advection}), and inverted there through the diagonalization process to
compute the fluxes $F^{n+1/2}_{i+1/2}$ that we will need in the second
step (\ref{steptwo}).

There are many ways to propagate the diagonal fluxes. Let us consider for
instance a diagonal flux $F$ with positive characteristic speed $\lambda$. Let
us decompose the resulting interface flux as follows:
\begin{equation}
F^{n+1/2}_{i+1/2} = F^n_i + \delta_i \;,
\label{delta}
\end{equation}
so that $F^n_i$ would be the zero order prediction and $\delta$ gives then the
difference between the zero and first or higher order estimates. Different
$\delta$ values will lead to different methods, like Lax-Wendroff,
\begin{equation}
\delta^{LW}_i = 1/2\;(1-\lambda\;{\Delta t \over \Delta x}) (F^n_{i+1} -
F^n_i)
\end{equation}
or Beam-Warning,
\begin{equation}
\delta^{BW}_i = \delta^{LW}_{i-1}
	      = 1/2\;(1-\lambda\;{\Delta t \over \Delta x}) (F^n_i - F^n_{i-1})
\end{equation}
and we see that, up to a factor, $\delta_i$ is a measure of the
flux slope at the upstream grid point $x_j$.

One can choose a slope which is a non-linear average of the two one-sided
linear estimates, namely
\begin{equation}
\delta_i = minmod(\delta^{LW}_{i-1}\;,\delta^{LW}_i) \; .
\label{average}
\end{equation}
which corresponds to the ``minmod'' average (just discard the steepest slope).
This leads to the ``Monotonic Upstream-Centered'' (MUSCL) algorithm, which is
very robust and easy to implement. Unlike the Lax-Wendroff or Beam-Warning
methods, it has the TVD property\cite{Leveque}, which means that no spurious
numerical oscillations can appear during the transport step, and this proves
to be very convenient in black hole codes, where large gradients arise near
the horizon.  We will show examples in a future publication.

\subsection*{Boundary Conditions}
The main problems with boundary conditions are consistency and stability.
Hyperbolic systems show here one of their main advantages, because they allow a
complete analysis of the information flow at the boundary. One can actually
compute the characteristic fields at every boundary and separate them into two
sets: either incoming or outgoing ones. Boundary conditions should provide
external information for incoming fields only. Imposing any condition on
outgoing fields would be inconsistent and boundary instability would arise in
the code. In this sense, an ``outgoing radiation'' condition should be
understood and implemented as a ``no incoming radiation'' condition for
the incoming fields.

The second order accurate operator splitting (\ref{splitting}) that we are
using requires boundary conditions to be imposed on the transport step only
(the sources step is made of ordinary differential equations). This is why we
have chosen the ordering given in (\ref{splitting}) instead of
\begin{equation}
E(\Delta t) = T(\Delta t/2)\;S(\Delta t)\;T(\Delta t/2) \;\; ,
\end{equation}
which would require to impose boundary conditions twice on a single timestep.

The natural way to impose boundary conditions in the two step schemes
(\ref{steptwo}) that we are using for the transport part is by prescribing the
fluxes $F^{n+1/2}$ of the incoming fields at the outermost interfaces
\begin{equation}
F^{n+1/2}_{imin-1/2} \;\;\; ,F^{n+1/2}_{imax+1/2}\; ,
\label{boundaries}
\end{equation}
where $imin$, $imax$ are the labels of the boundary points.

We have seen in the previous sections that, for every space direction,
there are six incoming eigenfields at the outer boundaries (five on the
light cone plus one on the gauge cone).
Remember, however, that we have three ``redundant'' quantities
$V_i$ in our formalism, which are related with the metric derivatives
through the algebraic equation (\ref{vector}), which amounts to the
momentum constraint.  This allows us to use the computed values of the
$V_i$ to reduce the number of algebraically independent incoming
fields at the boundary to what one could expect from physical
considerations: one gauge field plus two light cone fields which
account for the gravitational radiation degrees of freedom.

One could argue that the use of the constraint (\ref{vector}) at the
boundaries is not consistent in the operator splitting approach,
because the constraints are not first integrals of the transport and
source parts, when considered separately.  This is true in the generic
case, but one should notice that in the Strang splitting we are using
(\ref{splitting}) we impose the boundary conditions at $t^{n+1/2}$ in
the transport step, so that we have evolved precisely
\begin{equation}
T(\Delta t/2)\;S(\Delta t/2)
\end{equation}
and the values $u^{n+1/2}$ provide a consistent first order approximation of
the complete equation (source terms included) at $t^{n+1/2}$.

\section{Conclusions}
\label{sec:conclusion}
We have extended the discussion of our previous Letter on a new
hyperbolic formulation of Einstein's equations for numerical
relativity.  The equations have been cast into a flux conservative
system of balance laws that is valid for {\em any} choice of lapse
and shift.  This formulation allows one to use numerical methods such
as MacCormack, Lax-Wendroff, and others.  We further showed that the
system of equations is {\em hyperbolic}, meaning that its
characteristic matrix is fully diagonalizable with real eigenvalues,
for a wide variety of lapse conditions, including most of those
commonly used in numerical relativity.  This hyperbolicity can be
maintained for arbitrary shift choices, provided that care is taken in
the numerical implementation during the transport step of the
evolution scheme.

This hyperbolicity of the system can provide many
additional benefits, including the application of a large body of
numerical methods that make use of the eigenfields and their
characteristic speeds.  Such methods have been crucial in
hydrodynamics, where large gradients develop, and should find
application in numerical relativity as well.  We sketched in a general
way how some of these methods can be applied to our system of
equations.  Hyperbolicity can also provide information about the
fields, their direction of propagation, and their speeds, that should
prove to be very useful in devising boundary conditions, both at
horizons (AHBC) and at numerical grid boundaries.  In both cases, one
wants to enforce that outgoing signals leave the boundary, and that
no spurious signals come in.

Finally, we showed that our system of equations can be modified by use
of the hamiltonian constraint to remove numerically difficult terms in
the evolution equations, while maintaining hyperbolicity.  This leads
us to propose use of what we refer to as the ``Einstein'' system, as
opposed to the standard ``Ricci'' system.

Detailed numerical examples of these ideas will be provided in future
papers in this series.

\section*{Acknowledgments} We have benefited from discussions with a
number of our colleagues, particularly Wai-Mo Suen.  This work is
supported by the Direcci\'on General para la Investigaci\'on
Cient\'{\i}fica y T\'ecnica of Spain under project PB94-1177, and
also by NSF grants PHY94-07882 and INT94-14185.

\bibliographystyle{prsty}

\end{document}